\documentclass[aps,preprint,epsfig,rotate]{revtex4}
\usepackage{graphicx}
\usepackage{bm}
\usepackage{epsfig}

\input epsf
\begin{document}
\title{Stability and hyperfine structure of the four- and five-body 
       muon-atomic clusters $a^{+} b^{+} \mu^{-} e^{-}$ and 
       $a^{+} b^{+} \mu^{-} e^{-} e^{-}$.}

 \author{Alexei M. Frolov}
 \email[E--mail address: ]{afrolov@uwo.ca}

  \author{David M. Wardlaw}
  \email[E--mail address: ]{dwardlaw@uwo.ca}

\affiliation{Department of Chemistry\\
 University of Western Ontario, London, Ontario N6H 5B7, Canada}

\date{\today}

\begin{abstract}

Based on the results of accurate variational calculations we demonstrate stability 
of the five-body negatively charged ions $a^{+} b^{+} \mu^{-} e^{-} e^{-}$. Each 
of these five-body ions contains two electrons $e^{-}$, one negatively charged muon 
$\mu^{-}$ and two nuclei of the hydrogen isotopes $a, b = (p, d, t)$. The bound 
state properties of these five-body ions, including their hyperfine structure, are 
briefly discussed. We also investigate the hyperfine structure of the ground states 
of the four-body muonic quasi-atoms $a^{+} b^{+} \mu^{-} e^{-}$. In particular, we 
determine the hyperfine structure splittings for the ground state of the four-body 
muonic quasi-atoms: $p^{+} d^{+} \mu^{-} e^{-}$ and $p^{+} t^{+} \mu^{-} e^{-}$.

\end{abstract}

\maketitle
\newpage

\section{Introduction}

In our earlier work \cite{FrWa2011} we have discussed the four-body muonic quasi-atoms 
$a^{+} b^{+} \mu^{-} e^{-}$ and some of their properties. Here and everywhere below, 
the notations $a$ and $b$ stand for the nuclei of the hydrogen isoptopes ($p$ designates
protium, $d$ is deuterium and $t$ is tritium), $\mu^{-}$ (or $\mu$) and $e^{-}$ 
(or $e$) designate the negatively charged muon and  electron, respectively. These muonic
quasi-atoms are formed in liquid hydrogen/deuterium during muon-catalyzed fusion. It 
appears that the four-body muonic quasi-atoms, e.g, $p^{+} d^{+} \mu^{-} e^{-}, d^{+} 
d^{+} \mu^{-} e^{-}$, etc, have many extraordinary properties which follow from their
two-shell cluster structure. In particular, each of these ions has a heavy central 
nucleus $a b \mu$ which is positively charged. Spatial radius of such a three-body
nucleus is $R_{\mu} \approx \frac{a_0}{m_{\mu}}$. One electron moves around this heavy 
nucleus at disctances $R_e \approx a_0$, where $a_0$ is the Bohr radius. It is clear 
that $R_e \gg R_{\mu}$, since $\frac{m_{\mu}}{m_e} \approx 206.768$. 

The two-shell cluster structure of these quasi-atoms allows one to predict a large 
number of their bound state properties without extensive numerical calculations. To
check the overall accuracy of such predictions in \cite{FrWa2011} we have performed 
variational computations of the ground states in the six muonic atoms: $p p \mu e, p d 
\mu e, p t \mu e, d d \mu e, d t \mu e$ and $t t \mu e$. In such calculations we used 
the following particle masses $m_p$ = 1836.152701 $m_e$, $m_d$ = 3670.483014 $m_e$, 
$m_t$ = 5496.92158 $m_e$ and $m_{\mu}$ = 206.768262 $m_e$. These values of the particle 
masses are often used in recent computations of muonic molecular ions $a b \mu$ and 
other related systems. The same masses will be used in this study. 

In this study we continue our analysis of the four-body quasi-atoms $a^{+} b^{+} \mu^{-} 
e^{-}$ and closely related five-body negatively charged ions $a^{+} b^{+} \mu^{-} e^{-} 
e^{-}$ (or $a^{+} b^{+} \mu^{-} e^{-}_2$). As follows from our preliminary evaluations 
(see below) the correspoding five-body negatively charged ions $a^{+} b^{+} \mu^{-} e^{-} 
e^{-}$ are stable and can be observed in actual experiments. These ions are of great 
interest for general theory of bound states in few-body systems with unit charges. In this 
study we want to investigate the stability of these five-body ions formed by the particles 
with unit charges. Another aim is to determine the hyperfine structure splitting of the 
ground states in the five-body $a b \mu e_2$ ions and four-body $a b \mu e$ hydrogen-like 
quasi-atoms. In particular, we discuss the hyperfine structure and determine the 
hyperfine structure splitting for the ground state of the $p d \mu e$ quasi-atom and
predict the results of possible future experiements.  

\section{Bound states in the four-body muonic quasi-atoms}

In our variational calculations of the four-body quasi-atoms $a^{+} b^{+} \mu^{-} e^{-}$ 
in \cite{FrWa2011} we have applied the variational expansion written in the basis of the 
six-dimensional (or four-body) gaussoids of relative (or inter-particle) scalar coordinates 
$r_{ij}$. This variational expansion was originally proposed thirty years ago in \cite{KT} 
for accurate variational calculations of few-nucleon nuclei. For the bound $S(L = 
0)-$states the variational anzatz of six-dimensional (or four-body) gaussoids in relative 
coordinates is written in the form \cite{KT}
\begin{eqnarray}
 \Psi_{L=0} = \frac{1}{\sqrt{2}} (1 + \delta_{ab} \epsilon_{ab} {\cal P}_{ab}) \sum_{k=1}^N
 C_k \cdot exp( -\alpha^{(k)}_{12} r^{2}_{12} -\alpha^{(k)}_{13} r^{2}_{13}
 -\alpha^{(k)}_{23} r^{2}_{23} -\alpha^{(k)}_{14} r^{2}_{14}
 -\alpha^{(k)}_{24} r^{2}_{24} -\alpha^{(k)}_{34} r^{2}_{34}) \label{Gaus}
\end{eqnarray}
where $C_k$ are the linear coefficients (or linear variational parameters), while 
$\alpha^{(k)}_{ij}$ are the optimized non-linear parameters. The notation $\epsilon_{ab} 
{\cal P}_{ab}$ means the appropriate symmetrizer (or antisymmetrizer), i.e. a projection 
operator which produces the wave function with the correct permutation symmetry in those 
cases when $a = b$. This case is designated in Eq.(\ref{Gaus}) with the use of the
delta-function. The operator ${\cal P}_{ab}$ is the pair-permutation operator for all 
coordinates, i.e. for the spatial, spin, iso-spin, etc, coordinates. In these calculations
we have used two trial wave functions with $N$ = 400 and $N$ = 600 terms, respectively.
All non-linear parameters $\alpha^{(k)}_{ij}$ in Eq.(\ref{Gaus}) have been varied carefully.

The results of new variational computations of the ground states in the six muonic quasi-atoms
$p p \mu e, p d \mu e, p t \mu e, d d \mu e, d t \mu e$ and $t t \mu e$ can be found in 
Table I, where the total energies of these quasi-atoms are given in atomic units 
$\hbar = 1, m_e = 1, e = 1$. These energies are slightly better than the analogous total 
energies obtained in \cite{FrWa2011}. In this study we need the improved total energies of 
the four-body muonic quasi-atoms $a^{+} b^{+} \mu^{-} e^{-}$ to show the stability of the 
corresponding five-body ions $a^{+} b^{+} \mu^{-} e^{-} e^{-}$ which contain two electrons 
(see below). Note also that there are twenty two bound states in the six three-body muonic 
molecular ions $p p \mu, p d \mu, p t \mu, d d \mu, d t \mu$ and $t t \mu$. Formally, each
of these ions (in their ground and/or excited states) can bind an electron and form the 
four-body quasi-atom $a b \mu e$. In this study we restrict ourselves to the analysis of the 
six ground states only. 
 
\section{Stability of the  five-body muonic ions}

As follows from the general theory, the stability of the four-body quasi-atoms $a^{+} b^{+} 
\mu^{-} e^{-}$ means that the corresponding two-electron ions $a^{+} b^{+} \mu^{-} e^{-} 
e^{-}$ (or $a^{+} b^{+} \mu^{-} e^{-}_2$) are also stable. Each of these ions must have only 
one bound state which is, in fact, a weakly bound state, i.e. its binding energy is $\le$ 1 
\% of its total energy. Here, the stability means the energy stability against the following 
two-particle dissociation into quasi-atom $a b \mu e$ and free electron 
\begin{equation}
 a^{+} b^{+} \mu^{-} e^{-} e^{-} = a^{+} b^{+} \mu^{-} e^{-} + e^{-} \label{eq1}
\end{equation}
Briefly, the word `stability' means that the total energy of the five-body muonic ion $a^{+} 
b^{+} \mu^{-} e^{-} e^{-}$ is lower than the total energy of the four-body $a^{+} b^{+} 
\mu^{-} e^{-}$ quasi-atom in its ground state. Furthermore, the general theory of bound 
states in few-body systems predicts that the five-body $a^{+} b^{+} \mu^{-} e^{-} e^{-}$ ion 
has the only one bound $S(L = 0)-$state and this state is a singlet electron state. Formally, 
such a state can be designated as the $1^{1}S_e-$state, but in some cases this notation can 
lead to a confusion. In this Section we want to prove that the five-body negatively charged 
muonic ions $a^{+} b^{+} \mu^{-} e^{-} e^{-}$ are stable by performing accurate variational 
computations. 

First, note that the bound (ground) state of the $a^{+} b^{+} \mu^{-} e^{-} e^{-}$ ion
has the two-shell cluster structure. The internal three-body system, i.e. nucleus with
electric charge +1, is formed by the three heavy particles $a, b$ and $\mu$. The spatial 
radius of such a `quasi-nucleus' is $R_{\mu} \approx \frac{a_0}{m_{\mu}}$, where 
$m_{\mu}$ is the muon mass and $a_0$ is the Bohr radius. Everywhere below in this study we 
shall use only atomic units. In these units $\hbar = 1, e = 1$ and $m_e = 1$ and one finds 
$m_{\mu} = 206.768 262$ and $a_0 = 1$. This means that  $R_{\mu} \approx \frac{a_0}{206.768262}
\ll a_0$. The two electrons are moving around the central, positively charged 
`quasi-nucleus' at the distances $R_e \approx a_0 = 1$. In the zero-order approximation we 
can assume that $R_{\mu} = 0$ and $R_e = 1$. This leads us to the hydrogen-like 
quasi-atoms $a b \mu e$ and negatively charged ions $a b \mu e e$. However, even in this
approximation we cannot ignore correlations between different particles, and first of all 
electron-muon correlations. In the next order approximation when $R_{\mu} > 0$ the 
situation becomes more complicated. This requires an additional flexibility of
the variational anzatz which is used to construct trial wave functions. In all five-body 
calculations performed in this study we have used the variational anzatz which is written in 
the ten-dimensional gaussoid functions (or five-body gaussoids) of the ten relative coordinates 
$r_{ij}$ \cite{KT}
\begin{eqnarray}
 \Psi_{L=0} = (1 + \delta_{ab} \epsilon_{ab} {\cal P}_{ab}) 
 (1 + {\cal P}_{45}) \sum_{k=1}^N\sum_{(ij)=(12)}^{(45)}
 C_k \cdot exp( -\alpha^{(k)}_{ij} r^{2}_{ij}) \label{Gaus}
\end{eqnarray}
where $C_k$ are the linear coefficients (or linear variational parameters), while 
$\alpha^{(k)}_{ij} (= \alpha_{ji}$) are the optimized non-linear parameters. It is assumed 
in Eq.(\ref{Gaus}) that $i < j$. The notation ${\cal P}_{ab}$ means the permutation 
operator of the two particles $a$ (particle 1) and $b$ (particle 2). Additional phase 
$\delta_{ab} \epsilon_{ab} = \pm \delta_{ab}$, where $\delta_{km}$ is the Kronecker delta, 
is needed to provide the correct permutation symmetry in those systems where $a = b$. 
The analogous operator ${\cal P}_{45}$ is a pair permutation of the two identical electrons 
(particles 4 and 5) in the five-body $a b \mu e e$ ion. The overall electron symmetry of the 
trial wave function, Eq.(\ref{Gaus}), corresponds to the singlet electron state(s).

The results of numerical calculations of the ground state energies of the symmetric five-body
ions $p^{+} p^{+} \mu^{-} e^{-} e^{-}, d^{+} d^{+} \mu^{-} e^{-} e^{-}$ and  
$t^{+} t^{+} \mu^{-} e^{-} e^{-}$ are shown in Table I. In these calculations we have used
400 basis functions in Eq.(\ref{Gaus}), i.e. $N = 400$. The same Table contains the total 
variational energies of the ground states of the symmetric four-body quasi-atoms: $p^{+} 
p^{+} \mu^{-} e^{-}, d^{+} d^{+} \mu^{-} e^{-}$ and  $t^{+} t^{+} \mu^{-} e^{-}$ (see 
Section II). As follows from Table I the total energies of the five-body ions $p^{+} p^{+} 
\mu^{-} e^{-} e^{-}, d^{+} d^{+} \mu^{-} e^{-} e^{-}$ and $t^{+} t^{+} \mu^{-} e^{-} e^{-}$ 
are lower, i.e. they are more negative than the total energies of the corresponding four-body 
quasi-atoms $p^{+} p^{+} \mu^{-} e^{-}, d^{+} d^{+} \mu^{-} e^{-}$ and  $t^{+} t^{+} \mu^{-} 
e^{-}$. Note again that in our four-body computations we have used the trial wave functions 
with $N$ = 400 and 600 terms and all total energies of the five-body ions have been 
determined with the use of $N$ = 400 basis functions, Eq.(\ref{Gaus}). This means the absolute 
stability of the five-body ions $a^{+} a^{+} \mu^{-} e^{-} e^{-}$ against dissociation as 
represented by Eq.(\ref{eq1}). By performig analogous calculations for the corresponding five- 
and four-body non-symmetric ions $a^{+} b^{+} \mu^{-} e^{-} e^{-}$ and quasi-atoms $a^{+} 
b^{+} \mu^{-} e^{-}$ one finds that these ions are also stable.

Let us compare our results obtained for the five-body $a^{+} b^{+} \mu^{-} e^{-} e^{-}$
ions with the known total energies of the hydrogen atom H and hydrogen ion H$^{-}$ with the 
infinitely heavy nucleus. Traditionally in atomic physics such systems are designated as the 
${}^{\infty}$H atom and ${}^{\infty}$H$^{-}$ ion, respectively. The best-to-date variational 
energy obtained for the hydrogen ${}^{\infty}$H$^{-}$ ion \cite{Fro07H} is -0.527 751 016 
544 377 196 589 733 $a.u.$, where 21 - 22 decimal digits are stable. The total energy of the 
non-relativistic ${}^{\infty}$H atom is -0.5 $a.u.$ (exactly). Therefore, the corresponding 
binding energy of the negatively charged hydrogen ion is $\approx$ -0.027751 $a.u.$ 
Approximately the same binding energy is obtained for each of the five-body ions $a^{+} a^{+} 
\mu^{-} e^{-} e^{-}$ considered here. In reality, one finds the two following differences 
between the hydrogen ${}^{\infty}$H$^{-}$ ion and five-body $a^{+} b^{+} \mu^{-} e^{-} e^{-}$ 
ion: (1) the mass of the central quasi-nucleus $a^{+} b^{+} \mu^{-}$ is always finite, and (2) 
there are correlations between electrons and heavy particles from the central quasi-nucleus. 
It is clear that the electron-muon correlations play the leading role here. In general, the 
contribution of such correlations slightly decreases the actual binding energy of the five-body 
$a^{+} b^{+} \mu^{-} e^{-} e^{-}$ ion.   

\section{Hyperfine structure of the ground states of the four-body quasi-atoms}

Now, consider the hyperfine structure splitting of the ground states in the four-body 
quasi-atoms $a^{+} b^{+} \mu^{-} e^{-}$ and five-body negatively charged ions 
$a^{+} b^{+} \mu^{-} e^{-} e^{-}$. Note, that the ground state of the five-body $a^{+} b^{+} 
\mu^{-} e^{-} e^{-}$ ion is the electron singlet state. The total electron spin of such a
state equals zero exactly. Therefore, the electron spin does not contribute to the overall
hyperfine structure. Briefly, we can say that the hyperfine structure of the five-body ion
$a^{+} b^{+} \mu^{-} e^{-} e^{-}$ exactly  coincides with the hyperfine structure of the
three-body central `quasi-nucleus' $a^{+} b^{+} \mu^{-}$. The hyperfine structures of the 
ground $S(L = 0)-$states in six muonic molecular ions $a^{+} b^{+} \mu^{-}$ have been 
analyzed recently in \cite{Hyper}. 

In contrast with the five-body (two-electron) ions $a^{+} b^{+} \mu^{-} e^{-} e^{-}$, the
corresponding four-body quasi-atoms have very interesting hyperfine structure. Below, in 
this study we restrict ourselves to the consideration of the $p^{+} d^{+} \mu^{-} e^{-}$ and
$p^{+} t^{+} \mu^{-} e^{-}$ quasi-atoms. The proton, triton, muon and electron have spin 
equal $\frac12$, while deuteron's spin equals 1. The dimension of the total spin-space for 
the $p d \mu e$ quasi-atom is $2 \times 3 \times 2 \times 2 = 24$. The representation of the 
rotation group acting in this spin space is represented as the direct sum of the corresponding 
irreducible representations. The ranks of such irreducible representations essentially 
determine the hyperfine structure, while the differences between hyperfine energy levels can 
be obtained by solving the corresponding eigenvalue problem (all details can be found in 
\cite{Fro07}). The Hamiltonian $(\Delta H)_{h.s.}$ which represents the spin-spin interactions 
takes the following form (in atomic units) (see, e.g., \cite{LLQ})
\begin{eqnarray}
 (\Delta H)_{h.s.} = \frac{2 \pi}{3} \alpha^2 \frac{g_p g_d}{m^2_p}
 \langle \delta({\bf r}_{pd}) \rangle ({\bf s}_p \cdot {\bf s}_d)+
 \frac{2 \pi}{3} \alpha^2 \frac{g_p g_{\mu}}{m_p m_{\mu}}
  \langle \delta({\bf r}_{p\mu}) \rangle ({\bf s}_p \cdot {\bf s}_{\mu}) 
 \nonumber \\
 + \frac{2 \pi}{3} \alpha^2 \frac{g_d g_{\mu}}{m_p m_{\mu}}
  \langle \delta({\bf r}_{d\mu}) \rangle ({\bf s}_d \cdot {\bf s}_{\mu})
 + \frac{2 \pi}{3} \alpha^2 \frac{g_p g_{e}}{m_p}
  \langle \delta({\bf r}_{p e}) \rangle ({\bf s}_p \cdot {\bf s}_{e}) \nonumber \\
  + \frac{2 \pi}{3} \alpha^2 \frac{g_d g_{e}}{m_p}
  \langle \delta({\bf r}_{d e}) \rangle ({\bf s}_d \cdot {\bf s}_{e}) 
  + \frac{2 \pi}{3} \alpha^2 \frac{g_{\mu} g_{e}}{m_{\mu}}
  \langle \delta({\bf r}_{\mu e}) \rangle ({\bf s}_{\mu} \cdot {\bf s}_{e}) 
 \label{eq3}
\end{eqnarray}
where $\alpha = \frac{e^2}{\hbar c}$ is the fine structure constant, $m_{\mu}$ 
and $m_p$ are the muon and proton masses, respectively. The factors $g_{\mu}, 
g_{p}, g_{d}$ and $g_{e}$ are the corresponding $g-$factors. The expression for
$(\Delta H)_{h.s.}$ is, in fact, an operator in the total spin space which has 
the dimension $(2 s_p + 1) (2 s_d + 1) (2 s_{\mu} + 1) (2 s_e + 1) = 24$. In our 
calculations we have used the following numerical values for the constants and 
factors from Eq.(\ref{eq3}): $\alpha = 7.297352586 \cdot 10^{-3}, m_p = 
1836.152701 m_e, m_{\mu} = 206.768262 m_e, g_e$ = -2.0023193043718 and $g_{\mu} 
= -2.0023218396$. The $g-$factors for the proton and deuteron are deteremined 
from the formulas: $g_p = \frac{{\cal M}_p}{I_p}$ and $g_d = \frac{{\cal 
M}_d}{I_d}$, where ${\cal M}_p = 2.792847386$ and ${\cal M}_d = 0.857438230$ are 
the magnetic moments (in nuclear magnetons) of the proton and deuteron, 
respectively. The spin of the proton and deuteron is designated in Eq.(\ref{eq3}) 
as $I_p = \frac12$ and $I_d = 1$.  
 
The energy levels of the hyperfine structure and the corresponding structure 
splittings (differences between energy levels) can be found in Table II. Note that 
these values are usually given in $MHz$, while the values of $(\Delta H)_{h.s.}$  
which follow from Eqs.(\ref{eq3})) are expressed in atomic units. To re-calculate 
them from atomic units to $MHz$ the conversion factor 6.57968392061 $\cdot 10^9$ 
$MHz/a.u.$ was used. The general hyperine structure of the ground state of the 
$p d \mu e$ quasi-atom follows from Table II. The total spin function has twenty
four components. It can be represented as a direct sum of the following irreducible 
components (from the top to the bottom): (1) $J = \frac32$ (four states), (2) $J =
\frac52$ (six states), (3) $J = \frac12$ (two states), (4) $J = \frac32$ (four states),
(5) $J = \frac12$ (two states), (6) $J = \frac32$ (four states) and (7) $J = \frac12$ 
(two states). The total final dimension of this direct sum is: 4 + 6 + 2 + 4 + 2 + 4 
+ 2 = 24 as expected. The hyperfine structure splittings, i.e. the differences $\Delta$,
are shown in Table II. There are three large and three small $\Delta$. In particular, 
the values of $\Delta_{12}, \Delta_{34}$ and $\Delta_{67}$ are small. Very likely these 
differences are related with the protium-deuterium spin-spin interaction which is very 
small by its absolute value, since the expectation value of the proton-deuteron 
delta-function is very small \cite{Hyper}. On the other hand, the values of 
$\Delta_{23}, \Delta_{45}$ and $\Delta_{56}$ are relativey large. They are related with 
the electron-nuclear, electron-muon and muon-nuclear spin-spin interactions. A slightly 
more contrast picture can be obtained for the ground states in other muonic quasi-atoms, 
e.g., $d t \mu e$ and $t t \mu e$. We have determined the hyperfine structures for the 
ground states in all six muonic quasi-atoms mentioned in \cite{FrWa2011}. The hyperfine
structure of the $p t \mu e$ quasi-atom can be found in Table III. In this case $g_t = 
\frac{{\cal M}_t}{I_t}$, where ${\cal M}_p = 2.792847386$ and $I_t = \frac12$. The 
hyperfine structure of the $p d \mu e, p t \mu e$ and other similar quasi-atoms which 
contain Coulomb three-body quasi-nuclei $a b \mu$ must be confirmed in future 
experiments.            

\section{Conclusion}

Thus, we have illustrated by numerical means the stability of the bound five-body ions 
$a b \mu e e$ in their ground states. Some of the properties of these five-body ions are 
also predicted. The hyperfine structure of these ions is identical to the hyperfine 
structure of the corresponing muonic molecular ions $a b \mu$. We also investigate the 
hyperfine structure of the four-body muonic quasi-atoms $a b \mu e$ which is 
substantially more complicated. To obtain this hyperfine structure splitting we have 
diagonalized the Hamiltonians of the spin-spin interactions which include six different 
terms and overall dimension up to 36 (for the $d d \mu e$ and $(d d \mu)^{*} e$ 
quasi-atoms). The hyperfine structure of the $p d \mu e$ and $p t \mu e$ quasi-atoms is 
investigated in details.

\newpage

\begin{table}[tbp]
   \caption{The total non-relativistic total energies $E$ of the ground
            $S(L = 0)-$states in the five-body ions $a a \mu e e$ and 
            four-body quasi-atoms $a a \mu e$ (in atomic units).}
     \begin{center}
     \begin{tabular}{llll}
        \hline\hline
      &  $p p \mu e e$  &  $d d \mu e e$  &  $t t \mu e e$  \\
        \hline
$E(N = 400)$ & -102.75052695 &  -110.34390610 & -113.49923563 \\
       \hline\hline
     &  $p p \mu e$  &  $d d \mu e$  &  $t t \mu e$  \\
        \hline
$E(N = 400)$ & -102.72331085 &  -110.31683802 & -113.47266365 \\

$E(N = 600)$ & -102.72332519 &  -110.31684129 & -113.47266986 \\
  \hline\hline
     &  $p d \mu e$  &  $p t \mu e$  &  $d t \mu e$  \\
        \hline
$E(N = 400)$ & -106.51226911 &  -107.99443693 & -111.86419287 \\

$E(N = 600)$ & -106.51233102 &  -107.99453043 & -111.86421475 \\
  \hline\hline
  \end{tabular}
  \end{center}
  \end{table}

\begin{table}[tbp]
    \caption{The hyperfine structure $E_{i}$ and hyperfine structure splitting 
             $\Delta_{i (i-1)}$ of the ground $S(L = 0)-$state of the $p d \mu 
             e$ quasi-atom (in $MHz$).}
      \begin{center}
      \begin{tabular}{lll}
        \hline\hline
         & $E_{i}$ $MHz$ & $\Delta_{i (i-1)}$ $MHz$ \\
      \hline
$\epsilon_{J=\frac32}$ &  1.2521503100$\cdot 10^7$ & ------------- \\

$\epsilon_{J=\frac52}$ &  1.2517912194$\cdot 10^7$ &  3.59090572$\cdot 10^3$ \\

$\epsilon_{J=\frac12}$ &  9.3086776002$\cdot 10^6$ &  3.209234594$\cdot 10^6$ \\

$\epsilon_{J=\frac32}$ &  9.3043876675$\cdot 10^6$ &  4.28993274$\cdot 10^3$ \\

$\epsilon_{J=\frac12}$ & -2.1222389818$\cdot 10^7$ &  3.052677749$\cdot 10^7$ \\

$\epsilon_{J=\frac32}$ & -2.3095865825$\cdot 10^7$ &  1.873476007$\cdot 10^6$ \\

$\epsilon_{J=\frac12}$ & -2.3100074249$\cdot 10^7$ &  4.208423455$\cdot 10^3$ \\
       \hline\hline
   \end{tabular}
   \end{center}
   \end{table}
\begin{table}[tbp]
    \caption{The hyperfine structure $E_{i}$ and hyperfine structure splitting 
             $\Delta_{i (i-1)}$ of the ground $S(L = 0)-$state of the 
             four-body $p t \mu e$ quasi-atom (in $MHz$).}
      \begin{center}
      \begin{tabular}{lll}
        \hline\hline
         & $E_{i}$ $MHz$ & $\Delta_{i (i-1)}$ $MHz$ \\
      \hline
$\epsilon_{J=1}$ &  1.5082272852$\cdot 10^7$ & ------------- \\

$\epsilon_{J=2}$ &  1.5078349188$\cdot 10^7$ &  3.9236642227$\cdot 10^3$ \\

$\epsilon_{J=0}$ &  1.6857006656$\cdot 10^6$ &  1.3392648522$\cdot 10^7$ \\

$\epsilon_{J=1}$ &  1.6776620651$\cdot 10^6$ &  8.0386005145$\cdot 10^3$ \\

$\epsilon_{J=1}$ & -3.1838284105$\cdot 10^7$ &  3.3515946170$\cdot 10^7$ \\

$\epsilon_{J=0}$ & -3.1842399042$\cdot 10^7$ &  4.1149362917$\cdot 10^3$ \\
       \hline\hline
   \end{tabular}
   \end{center}
   \end{table}
\end{document}